# Calibration of manganin pressure gauge for diamond-anvil cells

Jian Chen,[1] Hu Cheng,[1] Xuefeng Zhou,[1] Xiaozhi Yan,[1,2] Lingfei Wang,[1] Yusheng Zhao,[1,2] Shanmin Wang[1,2,*]

[1]Department of Physics, Southern University of Science and Technology, Shenzhen, Guangdong 518055, China

[2]SUSTech Academy for Advanced Interdisciplinary Studies & Shenzhen Engineering Research Center for Frontier Materials Synthesis at High Pressure, Southern University of Science and Technology, Shenzhen, 518055, China

**Abstract**

Pressure calibration for most diamond-anvil cell (DAC) experiments is mainly based on the ruby scale, which is key to implement this powerful tool for high-pressure study. However, the ruby scale can often hardly be used for programmably-controlled DAC devices, especially the piezoelectric-driving cells, where a continuous pressure calibration is required. In this work, we present an effective pressure gauge for DACs made of manganin metal, based on the four-probe resistivity measurements. Pressure dependence of its resistivity is well established and shows excellent linear relations in the 0 - 30 GPa pressure range with a slope of 23.4 (9) GPa for the first-cycle compression, in contrast to that of multiple-cycle compression and decompression having a nearly identical slope of 33.7 (4) GPa likely due to the strain effect. In addition, such-established manganin scale can be used for continuously monitoring the cell pressure of piezoelectric-driving DACs, and the reliability of this method is also verified by the fixed-point method with a Bi pressure standard. Realization of continuous pressure calibration for programmably-controlled DACs would offer many opportunities for study of dynamics, kinetics, and critical behaviors of pressure-induced phase transitions.

[*] Corresponding author. Email: wangsm@sustech.edu.cn

**Introduction**

Study of materials under pressure (P) over the past decades has discovered a large number of new materials and novel phenomena in condensed-matter physics[1]. These scientific findings are primarily achieved by exploiting high-P apparatus, including multi-anvil large-volume presses and DACs. Compared with most large-volume presses that are driven by oil pumps, in DAC devices the load is applied via mechanical means (e.g. via screws, either manually tightened or by means of a gear box), having a number of advantages[1,2], as they can conveniently be integrated with various state-of-the-art probes such as synchrotron radiation, neutron scattering, Raman scattering, Brillouin scattering and electrical transport measurements for high-P studies of phase transitions[1], metallic hydrogen[3-5], high-$T_c$ superconductive hydride[6-8], strongly-correlated systems[9], and so forth[10-14].

A major drawback of screw-driving DACs is related to their loading system, which does not allow a programmable control of external load, hence the cell pressure. As a result, pressures of most DAC experiments are manually controlled by tightening or loosing screws, resulting in step-by-step adjusted pressure conditions. In each pressure step, the cell pressure can only be post-determined by ruby scale or other optical methods[1,2], but the mechanical relaxations of cell body, cupped washers, gasket system, and pressure-transmitting media would produce unpredictable variation of pressure in the cell, which often leads to time-dependent effects on the measured pressure. As a result, the cell pressure often cannot be calibrated in real time (i.e., not 'in situ'), because most high-P experiments do not allow simultaneously obtain the signals from both sample and pressure standards.

Although the screw-driven DACs have been successfully used in certain experiments, especially where the continuous pressure control is not required, they can hardly be exploited for special measurements (e.g., electrical resistivity) that requires continuously varied pressure with rates ranging from 1 GPa/second to 10 GPa/millisecond in magnitude, hereafter defined as fast and ultrafast loading rates, respectively. This has posed an insurmountable technique barrier for study of some fundamental properties of materials at pressure, such as kinetics, dynamics, and critical behaviors of pressure-induced phase transitions[15, 16]. Note that the time scale for nucleation process of phase transition is in second or shorter as recently explored in ref. 17, and the DAC devices with fast pressure variation is highly desired. Thus, instrumental advancements along this line towards functionalizing DAC devices with accurate pressure control and in situ pressure calibration is in high demand for future high-P study.

To mitigate the shortages of conventional DAC devices, the membrane- and piezoelectric (PZT)-driving DACs have been recently developed[18-20], which hold great promises for a programmable control of cell pressure. By contrast to the membrane technique, the piezoelectric-driving system offers a more quantitative pressure control by applied voltage, in terms of the elongation of piezoelectric element on voltage[20, 21]. In addition, the loading rate of the cell can also be largely tuned from ultra-fast (~60 GPa/millisecond) to a very slow rate (~1 GPa/hour)[21], making it feasible for study of dynamics and kinetics of phase transitions[18, 22]. However, to date, the pressure of piezoelectric-driving DACs can only be calibrated by expensive ultra-fast spectroscopy using the ruby scale[21] or internal pressure standards using synchrotron radiation[23], which are unfavorable for applications in general laboratories, calling for a more suitable pressure gauge for this sort of DAC devices.

It has come to our attention that the electrical resistivity of some alloys is very sensitive to pressure, which may be used as pressure gauges for DAC experiments. Among them, manganin metal is one of the most promising candidates and has been used for pressure determination of clamp cells and large-volume apparatus[24, 25]. As a pressure standard, this material has a number of advantages. Firstly, manganin possesses a linear pressure-resistance variation up to 22 GPa[24, 26, 27], which was determined by the traditional two-probe method in large-volume presses. Secondly, in quasi-hydrostatic pressure conditions its resistivity change is expected to be reversible when pressure is released, which facilitates multi-cycle high-P experiments as demonstrated in large-volume press experiments[25]. Note that nonhydrostatic pressure may induce appreciable resistivity change as mentioned by Samara *et al.*[27]. Last, manganin has a nearly temperature-independent resistivity and it can also be used for low-temperature experiments without involving re-calibration of the gauge[28]. In fact, this material has been previously used as a pressure gauge for large-volume press on the basis of the relative resistance variation with pressure. But it cannot be applied for DAC experiments because of their small cell volumes, which limits the dimensions of involved manganin wire, leading to large experimental uncertainties in resistance. In contrast, the electrical resistivity is an intrinsic property of material and is independent on sample size, which can provide a more reliable pressure standard. However, the pressure-dependent resistivity of this material has not yet been well established.

In this work, we explore the possibility for calibrating cell pressure of DAC devices by the cheap manganin metal as a pressure standard, on the basis of electrical resistivity measurements. Using the four-probe method and ruby scale, pressure dependence of its resistivity has been well established. Based on the established method, we have successfully calibrated the pressure of piezoelectric-driving cells with external load that is programmably controlled.

**Experiment Details**

Manganin (i.e., Mn-Cu) wire with ~100 μm in diameter was used in the experiment and was composed of Cu: 84w%, Mn: 12w%, and Ni: 4w%. Prior to experiment, a manganin disk ~10 μm thick was prepared from a section of Mn-Cu wire and then loaded in the cell chamber of a DAC for electrical resistivity measurements, using the four-probe method. To establish the pressure-resistivity relations for manganin, we firstly performed the regular screw-driving DAC experiment and the cell pressure was calibrated by the rube scale[29] or Raman shift of anvil tip[30]. In these experiments, diamond anvils with 300-μm culet size were selected; the gasket system was made of stainless steel and cBN powders (~0.5 μm in grain size) glued by 10w% epoxy. The steel gasket was pre-indented and then machined by a laser drill to remove the indented area. The thus-formed sample hole was re-filled with epoxy-glued cBN powders and re-indented up to ~20 GPa. During this process, cBN/epoxy mixture deformed and flowed to spread over the surface of steal, hence formation of an insulating layer for electrically isolating both the metal samples and electrode wires from the steel gasket. On the other hand, cBN/epoxy also served as pressure-transmitting media to improve hydrostatic pressure conditions. A Keithley-2450 voltage meter was used as a DC current source and the voltage of sample was recorded by a DMM-7510-digit multimeter. Electrical resistance measurements were performed at a constant current of 100 mA. The resistivity data can be readily obtained by the well-known resistance (R)-resistivity (R') relation, R' = 2πRL, bridged by average distance (L) between the nearest neighboring electrodes. Before experiment, all four electrodes were intendedly mounted uniformly on the cutlet of diamond anvil with an equal electrode distance of L ≈ 50 μm. By optical observations, the electrode distances were monitored during compression, and the L value remained nearly invariant even at 30 GPa. Thus, the resistance (hence the resistivity) was almost independent on the sample dimension. A piezoelectric-driving DAC with a 500-μm culet size was employed for a programmable control of cell pressure by control of input voltage. A Bi pressure standard was also loaded in the same cell chamber with the manganin sample to check the accuracy of the calibrated manganin gauge. In principle, the electrical resistivity signal transmission by the manganin gauge using the four-probe method is as fast as the light speed, but the actual signal read-out rate is mainly dependent on the response of data collection system. Based on the current experimental setup, the highest sampling speed can reach 1 sample per 20 millisecond. But, using the high-speed data acquisition card, the data collection rate can be up to sub-millisecond level.

**Results and Discussion**

As shown in Fig. 1, we performed four-probe resistivity measurements for manganin using a conventional screwing-driving DAC. Pre-cutted Pt wires are used as electrodes and mounted on the culet of bottom diamond anvil. The tips of Pt wires should be sharp enough to avoid direct contacts between them during compression (Fig. 1a), because Pt is soft and will subject to a severe deformation under pressure. On the other hand, the soft Pt wires can withstand extraordinarily large shear deformation across the gasket area, in avoiding breakdown of the electrical circuit. It is also noted that the resistivity of Pt metal is very small and retains nearly invariant under deformation, which is an ideal electrode material for DAC experiments. As mentioned above, the steel gasket is covered by a thin epoxy-glued cBN, which can be used as an excellent insulating layer to electrically isolate the measurement circuit from steel gasket (Fig. 1b).

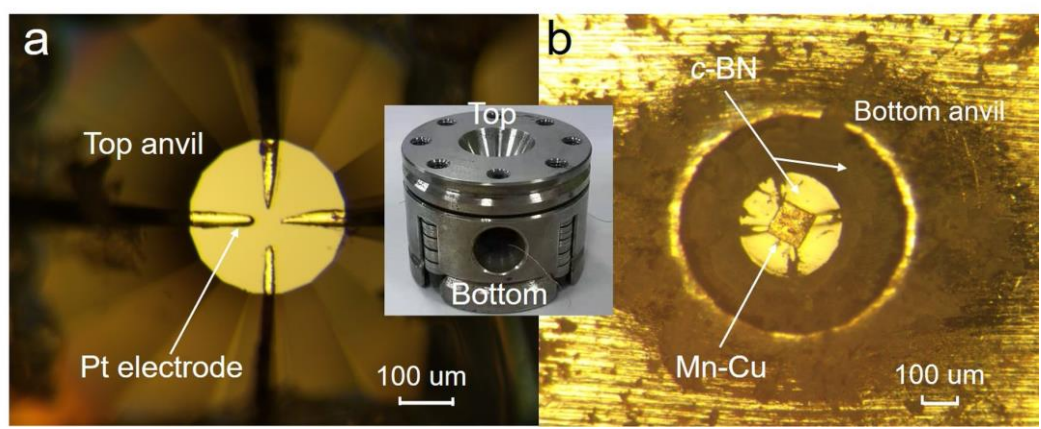

**Fig. 1.** Optical images of the sample cell with four electrodes made of Pt wires. (**a**) Four Pt electrodes mounted on the culet of the top diamond anvil. (**b**) A bird view of the prepared high-P chamber based on the bottom diamond anvil. In the center of the cell chamber, a Mn-Cu disk was loaded on surface of the compressed cBN/epoxy layer, which served as pressure-transmitting media. Four imprints on the surface of cBN/epoxy layer were formed by closing the two opposite anvils. Inset is an image of the DAC device for resistivity experiment.

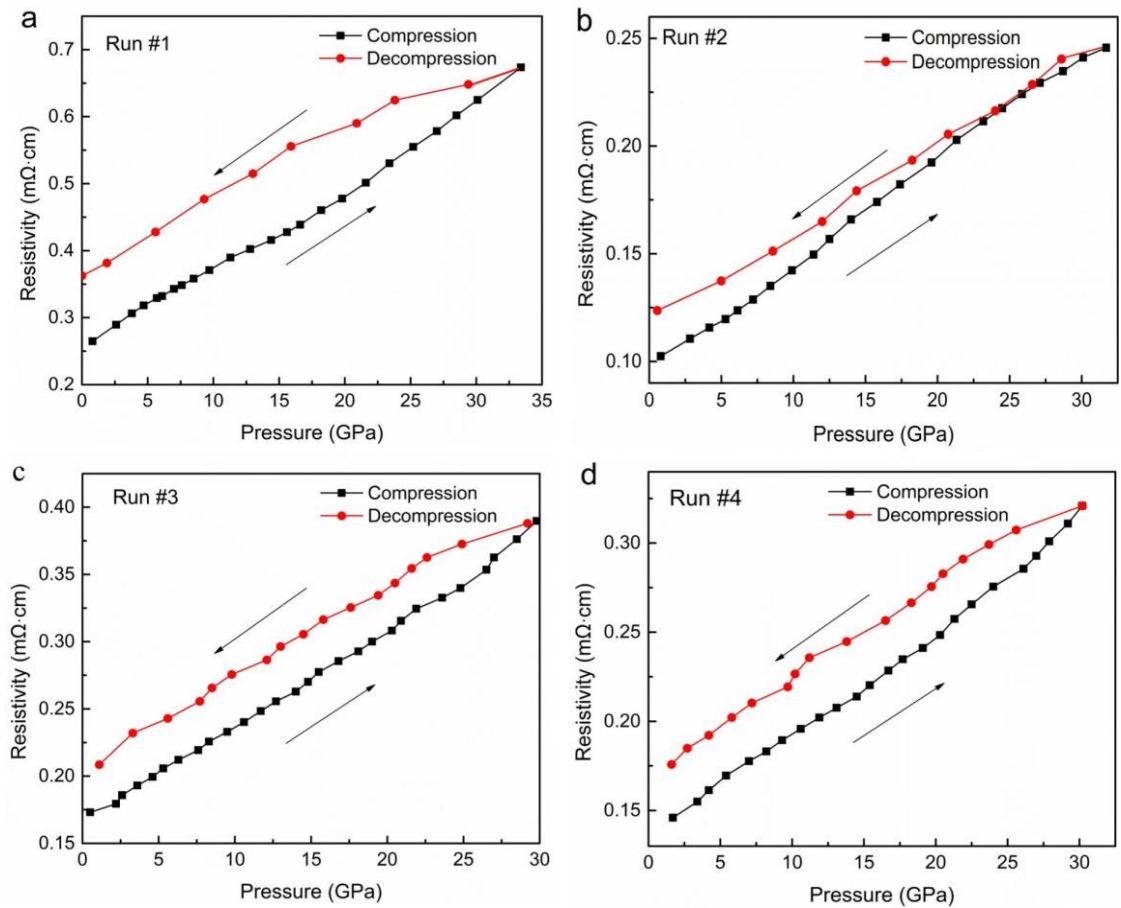

**Fig. 2.** Electrical resistivity of manganin as a function of pressure for four different experimental runs. (**a**) Run #1, (**b**) Run #2, (**c**) Run #3, (**d**) Run #4. Cell pressures of these experiments were determined by the ruby scale.

Fig. 2 shows the determined resistivity of Mn-Cu as a function of pressure and the experiments are repeated four times to obtain the statistical averages (see Figs. 2a-2d). Obviously, the resistivity has excellent linear relations with pressure up to 30 GPa for both compression and decompression. However, upon decompression the resistivity in each experimental run has a similar increase, when compared with that of compression, especially at the very beginning of decompression. This should be related to plastic deformation of manganin during compression, probably resulting from the nonhydrostatic pressure in the cell as previously reported[27]. Note that there have small resistivity differences between different experimental runs, probably due to the uncertainty in determining the distances between electrodes. To mitigate the influence of these resistivity differences, the relative resistivity ratio will be used for calibration of manganin gauge, as will be discussed later.

To quantitatively determine the linear pressure-resistivity relations, high-P resistivity ($R_p$) is normalized relative to the ambient value ($R_0$) and the resistivity ratio, $(R_p-R_0)/R_0$, is used for linear regression of the variation of resistivity against pressure, given by

$$P=K\times(R_p-R_0)/R_0. \qquad (1)$$

Where $R_0$ can readily be extrapolated from the pressure-resistivity curve, as seen in Fig. 3. $K$ represents the linear slope. In fact, the similar equation has been previously used for determining the pressure-resistance of manganin based on large-volume press experiments[24]. To have a comparison study, we plotted all datasets together in Fig. 3. Fits of the pressure-resistivity data to Eq. 1 yield detailed parameters and are summarized in Table 1. As expected, for each experiment run, the data taken on compression has a similar slope of $K_{Comp}$ = 23.4 (9) GPa (Fig. 3a), indicating a reliable slope value. By contrast, the decompression cases show a much better consistency with a nearly identical slope of $K_{Decomp}$ = 33.7 (4) GPa (Fig. 3b), which is ~40% larger than that of compression, as a result of plastic deformation of manganin at pressure. Remarkably, based on the obtained constant slopes of $K_{Comp}$ and $K_{Decomp}$, the pressure-resistivity relations of manganin gauge can be well established, which provide an effective approach for pressure determination of DAC devices. Also noted is that the involved experimental uncertainty by this gauge is very small and less than 0.5 GPa, especially in both the low-P range of compression (i.e., 0 - 10 GPa) and the whole decompression process. However, in the high-P range of compression (i.e., 10 - 30 GPa), the pressure uncertainty is increased, but it can still be limited within 2 GPa even at 30 GPa. Note that the resistance can also be used to do calibration for achieving the same accuracy, because the relative resistance ratio is identical to the relative resistivity ratio, giving rise to the same slope $K$ value, according to Eq. (1).

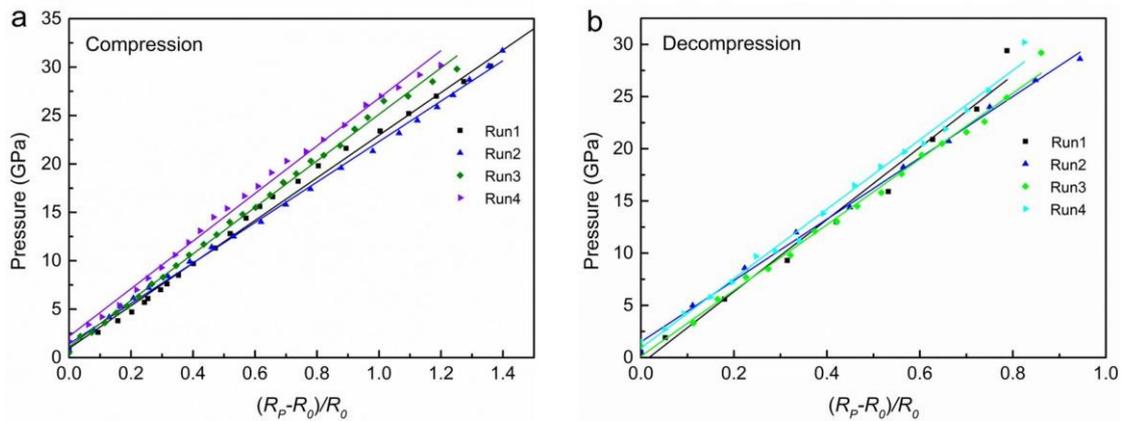

**Fig. 3.** Variation of relative resistivity ratio of manganin against pressure for different experimental runs. (**a**) Compression. (**b**) Decompression. Each solid line represents a linear fit of the corresponding pressure-dependent relative resistivity ratio curve.

**Table 1.** Derived linear slopes $K_{Comp}$ and $K_{Decomp}$ for four different experimental runs.

| Experiment | Compression | | Decompression | |
| --- | --- | --- | --- | --- |
| | $K_{Comp}$ (GPa) | Goodness of fit($R^2$) | $K_{Decomp}$ (GPa) | Goodness of fit($R^2$) |
| Run #1 | 22.7(4) | 0.9939 | 34.5(8) | 0.9874 |
| Run #2 | 22.2(9) | 09979 | 33.7(6) | 0.9965 |
| Run #3 | 24.0(3) | 0.9982 | 33.5(2) | 0.9977 |
| Run #4 | 24.6(4) | 0.9949 | 33.3(7) | 0.9937 |
| **Average** | 23.4(9) | | 33.7(4) | |

Pressure-induced deformation has great influence on the resistivity of manganin,

which accounts for the distinctly different slopes between compression and decompression as observed in Fig. 3. Multi-cycle compression-decompression processes may lead to accumulation of stress and plastic deformation, which would greatly change the slope, hence the pressure-resistivity relations. To check the multi-cycle effect on resistivity of manganin, we repeated the compression-decompression loading cycle for a few times in the experimental run #3 and #4, as shown in Fig. 4. Clearly, for the second cycle of run #3 (Figs. 4a-4b), either of its compression or decompression resistivity data have a similar trend to that of the first-cycle decompression data; the thus-determined slopes are 34.4 and 32.5 GPa, respectively (see Table 2), which are very close to that of the first-cycle decompression with a value of $K_{Decomp} \approx 33.5$ GPa. In the multi-cycle measurements of run #4, the similar results are also obtained as seen in Figs. 4c-4f, suggesting that the multi-cycle loading processes do not induce more plastic deformation or strain. It is evident that the obtained $K_{Decomp}$ can be applicable for pressure calibration of multi-cycle processes.

**Table 2** Derived linear slopes $K_{Comp}$ and $K_{Decomp}$ for the resistivity data taken on multi-cycle compression-decompression processes (see Fig. 4).

| Experiment | Cycle | Compression | | Decompression | |
| --- | --- | --- | --- | --- | --- |
| | | $K_{Comp}$ (GPa) | Goodness of fit(R2) | $K_{Decomp}$ (GPa) | Goodness of fit(R2) |
| Run #3 | Cycle1 | 24.0(3) | 0.9972 | 33.5(2) | 0.9917 |
| | Cycle2 | 34.4(7) | 0.9930 | 32.5(7) | 0.9932 |
| Run #4 | Cycle1 | 24.6(4) | 0.9949 | 33.3(7) | 0.9937 |
| | Cycle2 | 31.7(4) | 0.9985 | 34.9(7) | 0.9936 |

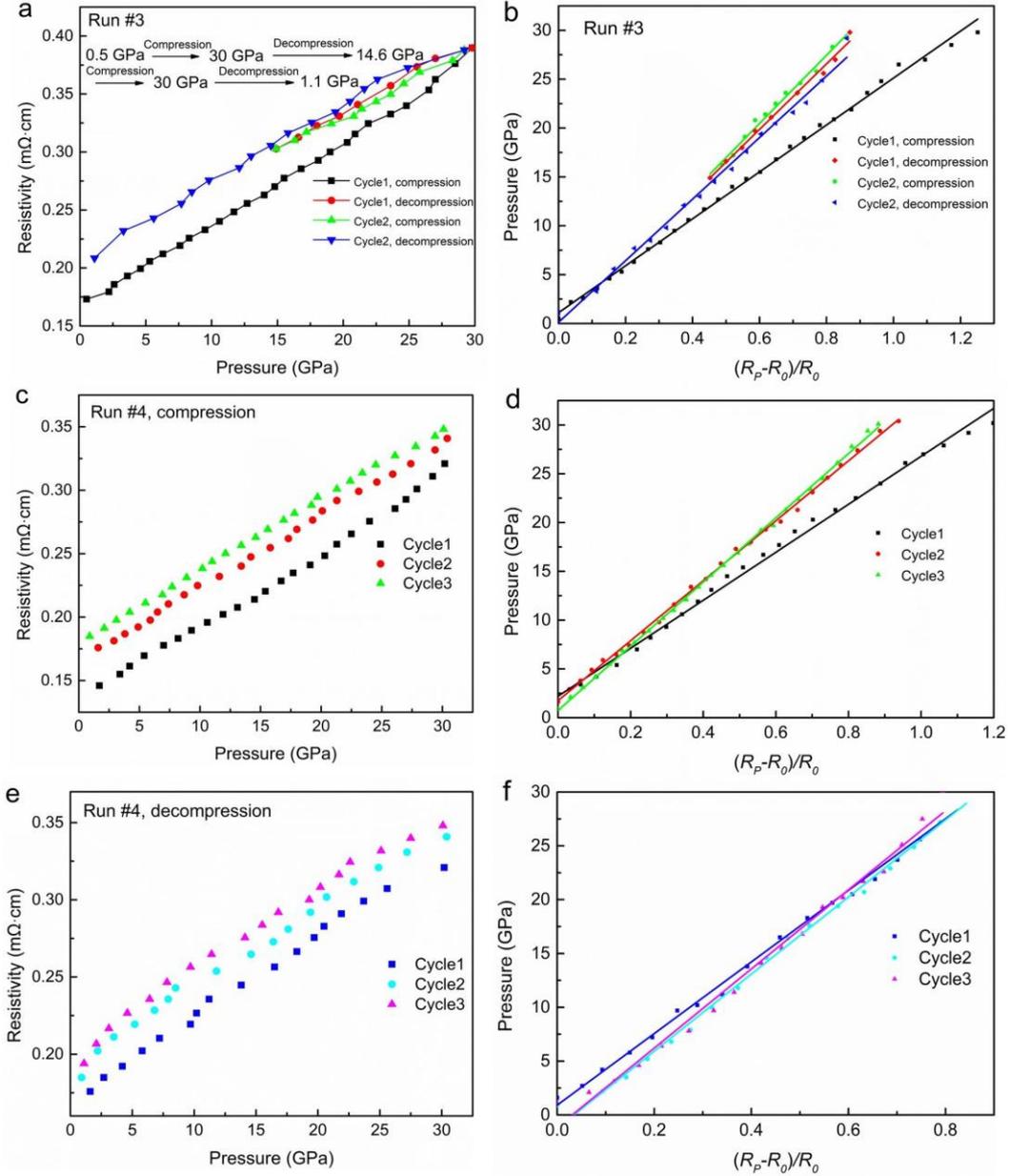

**Fig. 4.** Electrical resistivity of manganin vs. pressure for experimental run #3 and run #4 with resistivity data taken upon different compression-decompression loading cycles. (**a**) Resistivity vs. pressure for run #3. The path for compression-decompression loading cycles are listed in the figure. (**b**) Normalized resistivity ratio vs. pressure for run #3. (**c**) Resistivity vs. pressure taken upon compression for run #4. (**d**) Normalized resistivity ratio vs. pressure for (**c**). (**e**) Resistivity vs. pressure taken upon decompression for run #4. (**f**) Normalized resistivity ratio vs. pressure for (**e**). The solid lines in (b), (d), and (f) are linear regressions of the associated experimental data.

On the basis of resistivity measurements, the manganin gauge can readily realize fast and continuously monitoring pressure for piezoelectric- and membrane-driving DAC devices, without involvement of expensive ultrafast spectroscopy. This will be very useful for study of many fundamental issues (e.g., pressure-induced phase transitions). To check the possibility for applications of this gauge to the programmably-loaded cell, we performed the pressure calibration for a piezoelectric-driving DAC device using the manganin gauge, as shown in Fig. 5. The system is mainly composed of piezoelectric element, DAC, and steel sleeve (Figs. 5a-5b). The elongation of piezoelectric element is conveniently controlled by the control of input voltage, which drives the two opposite anvils of the DAC to squeeze the pressure-transmitting media and generate pressure in the cell. Note that the maximum elongation of the PZT element in our device is only 80 μm. Thus, it requires that the cell is pre-tightened with certain initial pressure to effectively use the elongation for obtaining higher pressure conditions. Because the response of elongation of piezoelectric element to the applied voltage is very short (i.e., in milliseconds), this device can be used for ultrafast pressure controls, which are important for investigating dynamics and kinetics of phase transitions at pressure. More detailed descriptions of piezoelectric-driving DAC devices can be found in previous reports[20, 21].

In order to verify the reliability of the calibrated manganin gauge, we also loaded a Bi pressure standard in the same cell chamber for resistivity measurement with a continuous variation of pressure, based on a piezoelectric-driving DAC. To facilitate preparation of enough electrodes (i.e., eight electrodes) for determining resistivity of two samples, diamond anvils with a culet size of as large as 500 μm are selected. Fig. 5c presents a thus-prepared cell with eight Pt electrodes; four electrodes are used for the manganin sample, while the rest four are connected with the Bi metal standard.

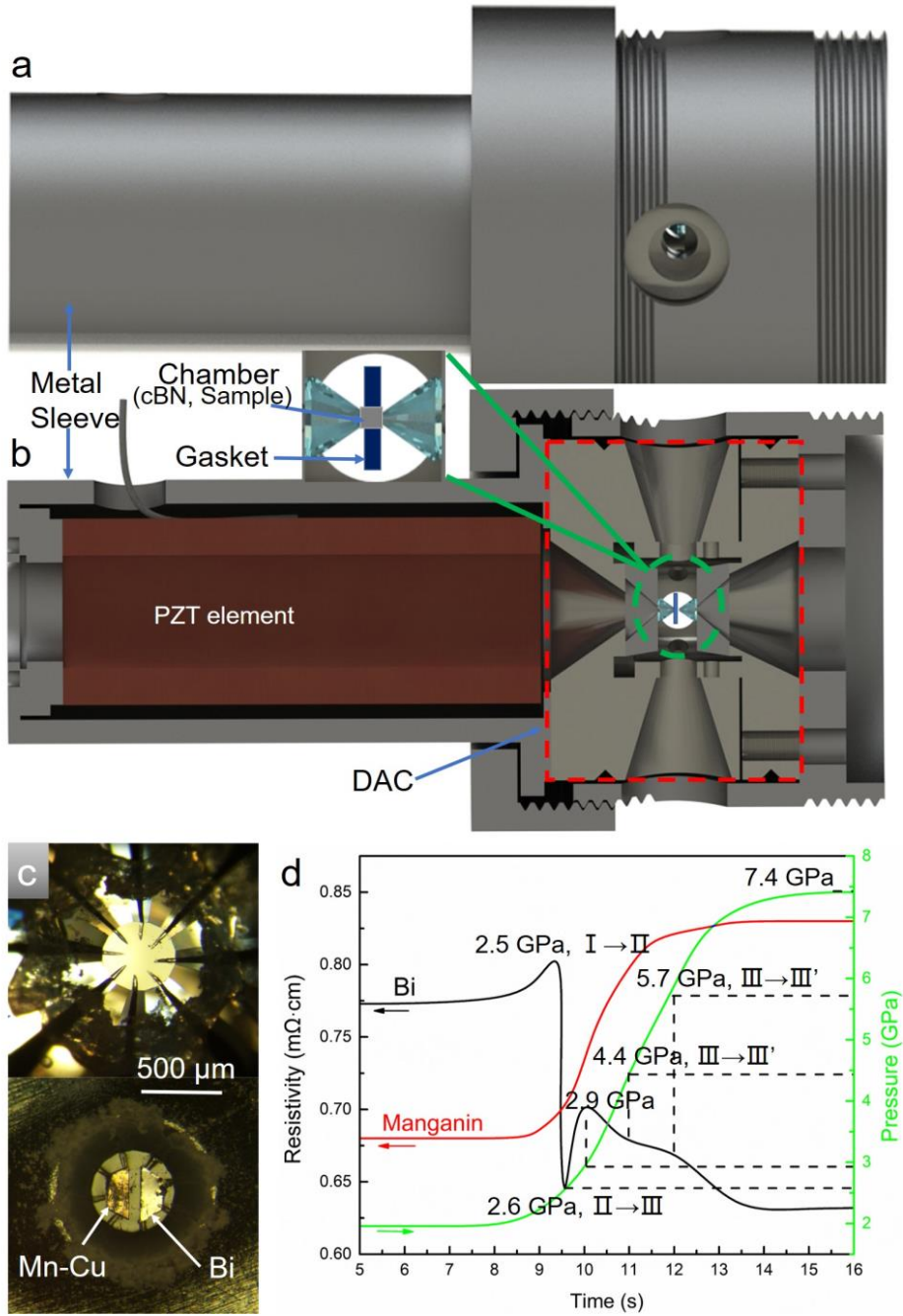

**Fig. 5.** Pressure calibration for a piezoelectric-driving DAC device using the calibrated manganin gauge. (**a**) Schematic diagram of piezoelectric-driving DAC with a sectional view in (**b**) to show details of internal structure. (**c**) Optical images of pre-prepared eight Pt electrodes on the culet of top anvil (upper panel). Four electrodes were used for manganin gauge and the rest four for Bi standard. Both samples were loaded on the surface of pre-compressed cBN/epoxy in the cell chamber based on the bottom anvil (lower panel). Eight imprints were formed by closing of two opposite diamond anvils. (**d**) Recorded resistivity of manganin (red line) and Bi (black line) vs. time. The green line represents the determined pressure curve using Eq. 1 with a slope of $K_{\text{Comp}} \approx 23.4$ GPa.

Fig. 5d shows the obtained resistivity as a function of pressure for both manganin and Bi metal. Using Eq. 1 and the pre-determined $K_{Comp}$ = 23.4 (9) GPa, the cell pressure can be easily calculated using the pressure-resistivity data of manganin. Intriguingly, the calculated pressure has a smooth variation with time, allowing a continuous, fast/ultrafast, and real-time pressure monitoring for the cell. For the Bi pressure standard, a number of resistivity changes start to occur at 2.5 GPa, 2.6 GPa, 2.9 GPa, 4.4 GPa, and 5.7 GPa, respectively, implying multiple phase transitions. The first two correspond to structural transitions of Bi: I-II (2.55 GPa) and II-III (2.69 GPa)[31-33], which are in excellent agreement with our calibrated values, substantiating the relibility of the established manganin scale. Due to the pressure-volume trade-off, the cell pressure is limited to 7.4 GPa because the diamond anvils used in this experiment has a large culet size (i.e., 500 μm), so that the well-known transition of Bi: III-V (7.7 GPa) with appreciable resistivity reduction is absent in our measurement (Fig. 5d). According to reported high-P phase diagram of Bi[31], the resistivity changes at 4.4 GPa and 5.7 GPa should correspond to the III → III' and III- III'' transitions; however, they have been rarely observed by conventional resistivity measurements based on large-volume pressures with a slow compression rate (i.e., less than 0.01 GPa/second) [34], likely because the associated phases are kinetically more unstable.

It is also strikingly noted that there seemingly has a new phase transition starts at 2.8 GPa (Fig. 5d), showing a rapid resistivity reduction, but no phase transition has been previously reported around this pressure based on large-volume press measurements. As mentioned above, the loading rate in this work is more than two orders of magnitude faster (i.e., with a rate of ~1 GPa/second) than that of conventional large-volume presses, which may allow recording the signals from this new metastable phase in a narrow pressure range. Future work along this direction is warranted to determine the crystal structures of this new phase of Bi metal on the basis of *in situ* fast spectroscopy. In addition, the loading rate of piezoelectric-driving DACs can extremely fast with a pressure rate up to ~60 GPa/millisecond[21], which, in combination of favorable pressure calibration method like manganin gauge, would provide great opportunities for study of dynamics of materials and discovery of new metastable phases at pressures. Besides, this established method is particularly favorable for high-P study of the samples that are photon degradable and therefore use of lasers is not advised, or if optical access is not immediately accessible as in some in-house cryostats not fitted with optical windows.

On a negative note, there are a few disadvantages associated with this method. Firstly, the major limitation of this method requires a very careful preparation of multiple electrodes by well-trained students. But recent developments of sputter coating and lithography techniques can help relax this experiment difficulty for ordinary users. Secondly, the multiple electrodes take large space, which would limit the pressure conditions because the involved diamond anvils have large culet size exceeding 300 μm. Nevertheless, with longer piezoelectric element the cell pressure can still go up to 30 GPa, which can satisfy the pressure requirements for most electrical transport experiments.

**Conclusions**

In summary, we have established a methodology for pressure calibration for DACs by four-probe resistivity measurements of the manganin metal. The linear pressure-resistivity relations have been well determined up to 30 GPa with two distinct slopes of $K_{\text{Comp}} \approx 23.4$ GPa and $K_{\text{Decomp}} \approx 33.7$ GPa for the first-cycle compression and multi-cycle processes, respectively. Using the calibrated manganin gauge, we have successfully determined the cell pressure of a piezoelectric-driving DAC with fast and continuous pressure variation in the cell. In addition, pressure-induced multiple phase transitions in Bi metal have also been checked by loading sample in the same cell chamber with manganin gauge, leading to discovery a new and a few sparsely reported metastable phases. Success in pressure calibration of piezoelectric-driving DACs will open a new avenue for exploring dynamics, kinetics, and critical phenomena in materials under pressure.

**Acknowledgments**

This work was supported by the Key Research Platforms and Research Projects of Universities in Guangdong Province (Grant No. 2018KZDXM062), the Guangdong Innovative & Entrepreneurial Research Team Program (No. 2016ZT06C279), the Shenzhen Peacock Plan (No. KQTD2016053019134356), the Shenzhen Development and Reform Commission Foundation for Shenzhen Engineering Research Center for Frontier Materials Synthesis at High Pressure, the Shenzhen Science and Technology Innovation Committee (Grant No. JCYJ20190809173213150), and the Research

Platform for Crystal Growth &Thin-Film Preparation at SUSTech.

**Author Information**

*E-mail: wangsm@sustech.edu.cn

**Notes**

The authors declare no competing financial interest.

**Data availability**

The data that support the findings of this study are available from the corresponding authors upon reasonable request.